\documentclass[twocolumn,aps,prl,superscriptaddress,preprintnumbers,bibnotes]{revtex4-2}

\usepackage[T1]{fontenc}

\usepackage{natbib}
\usepackage{graphicx}
\usepackage{bm}
\usepackage{color}
\usepackage{amsmath}
\usepackage{gensymb} 
\usepackage{physics}
\usepackage{tabularx}
\usepackage{hyperref}

\usepackage{changes}
\definechangesauthor[color=teal]{LP}
\definechangesauthor[color=magenta]{GEP}
\definechangesauthor[color=blue]{AJP}
\definechangesauthor[color=cyan]{JTZ}
\definechangesauthor[color=red]{KK}
\definechangesauthor[color=violet]{TR}

\newcommand{\Na}{Na}
\newcommand{\Cs}{Cs}
\newcommand{\CUAaddress}{Harvard-MIT Center for Ultracold Atoms, Cambridge, Massachusetts 02138, USA}
\newcommand{\HarvardPhysicsAddress}{Department of Physics, Harvard University, Cambridge, Massachusetts 02138, USA}
\newcommand{\HarvardChemistryaddress}{Department of Chemistry and Chemical Biology, Harvard University, Cambridge, Massachusetts 02138, USA}

\begin{document}

\title{Extended rotational coherence of polar molecules in an elliptically polarized trap}

\author{Annie J. Park}
\thanks{A.J.P. and L.R.B.P contributed equally to this work.}
\affiliation{\HarvardChemistryaddress} 
\affiliation{\HarvardPhysicsAddress}
\affiliation{\CUAaddress}

\author{Lewis R.B. Picard}
\thanks{A.J.P. and L.R.B.P contributed equally to this work.}
\affiliation{\HarvardPhysicsAddress}
\affiliation{\HarvardChemistryaddress} 
\affiliation{\CUAaddress}

\author{Gabriel E. Patenotte}
\affiliation{\HarvardPhysicsAddress}
\affiliation{\HarvardChemistryaddress} 
\affiliation{\CUAaddress}

\author{Jessie T. Zhang}
\altaffiliation[Current affiliation: ]{University of California, Berkeley}
\affiliation{\HarvardChemistryaddress} 
\affiliation{\HarvardPhysicsAddress}
\affiliation{\CUAaddress}

\author{Till Rosenband}
\affiliation{Agendile LLC, Cambridge, Massachusetts, 02140, USA}

\author{Kang-Kuen Ni}
\email{ni@chemistry.harvard.edu}
\affiliation{\HarvardChemistryaddress} 
\affiliation{\HarvardPhysicsAddress}
\affiliation{\CUAaddress}

\date{\today}

\begin{abstract}

We demonstrate long rotational coherence of individual polar molecules in the motional ground state of an optical trap.  In the present, previously unexplored regime, the rotational eigenstates of molecules are dominantly quantized by trapping light rather than static fields, and the main source of decoherence is differential light shift. In an optical tweezer array of \Na\Cs{} molecules, we achieve a three-orders-of-magnitude reduction in differential light shift by changing the trap's polarization from linear to a specific ``magic'' ellipticity. With spin-echo pulses, we measure a rotational coherence time of 62(3)~ms (one pulse) and 250(40)~ms (up to 72 pulses), surpassing the projected duration of resonant dipole-dipole entangling gates by orders of magnitude.
\end{abstract}

\maketitle

Protecting quantum systems from decoherence is necessary for quantum metrology, simulation, and information processing. Polar molecules are promising building blocks for these applications due to their rich identical structure, long coherence times  ~\cite{Park2017,Burchesky2021,gregory_robust_2021,Dajun2022}, and intrinsic anisotropic electric dipole-dipole interactions~\cite{yan_observation_2013,christakis_probing_2023}. Crucially, dipole-dipole interactions can deterministically entangle two rotation states of spatially separated molecules~\cite{demille_quantum02,ni_dipolar_2018}, as demonstrated recently with CaF molecules in optical tweezers~\cite{holland_-demand_2022, bao_dipolar_2022}. In such a case, where the molecules are directly laser cooled and loaded into optical tweezers, the fidelity of entanglement has been limited by residual thermal motion, which causes uncontrolled variation in the strength of entangling interactions.

This limit can be overcome using rovibrational-ground-state bi-alkali molecules prepared in the lowest motional state of an optical tweezer, for example by controlled association of ground-state-cooled individual atoms~\cite{cairncross_rovibgs_2021,zhang_optical_2022, ruttley2023formation}. The dominant source of decoherence is then due to the optical trap that spatially confines the molecules. The anisotropic polarizability of different rotational wavefunctions induces state dependent trap depths, leading to fluctuating transition frequencies and the dephasing of rotation states~\cite{Kotochigova2010}. Therefore, canceling differential light shifts is a major hurdle that must be overcome to achieve quantum coherence with these species as they undergo dipolar interactions.

\begin{figure}[t]
    \centering
\includegraphics[width=\columnwidth]{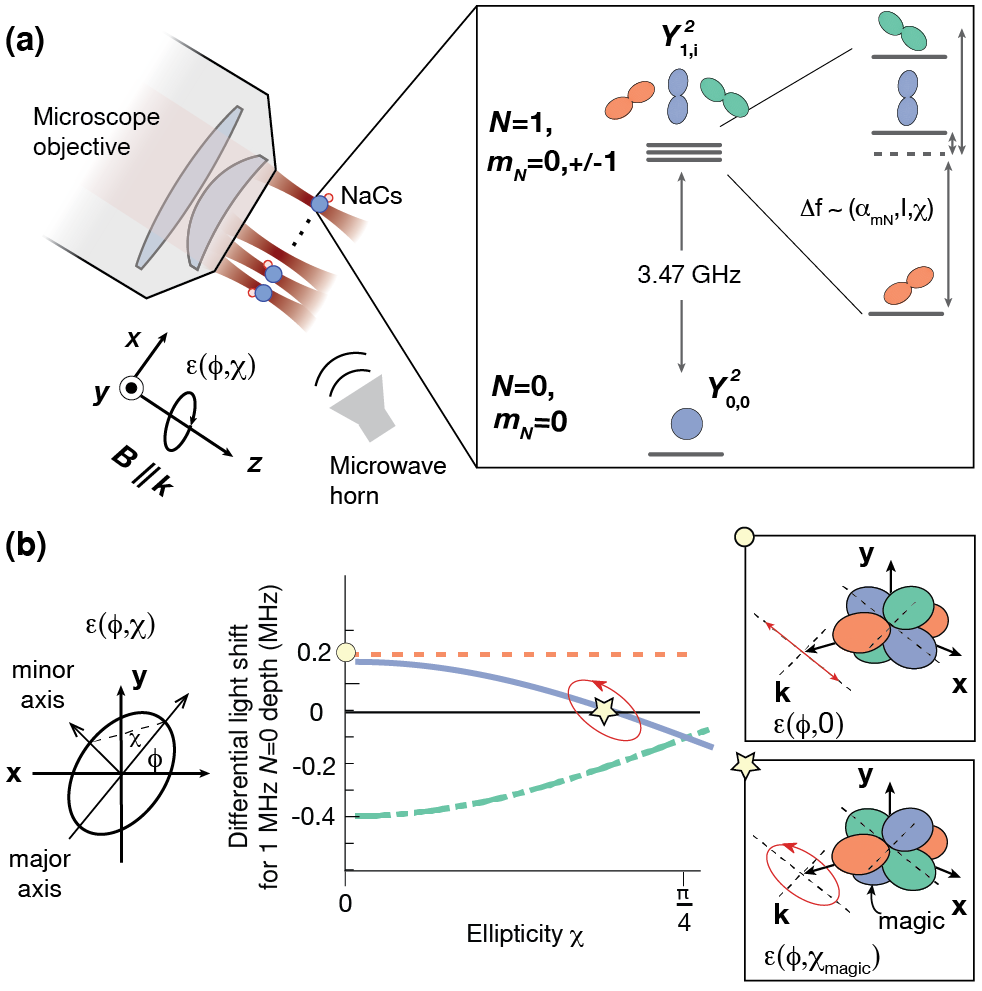}
    \caption{\Na\Cs{} molecules trapped in an optical tweezer array. (a) Schematic of the experimental setup, including the tweezer $\bm{k}$-vector, magnetic field $\bm{B}$, and trap polarization $\bm{\epsilon} (\phi,\chi)$.
     On the right is a simplified energy level diagram of ground ($N=0$) and first excited ($N=1$) rotational states, where the optical trap lifts the sublevel degeneracy.
    (b) The azimuthal angle $\phi$ and ellipticity $\chi$ of polarization determine the orientation and light shift respectively of the $N=1$ sublevels. Unlike for linear polarization (circle), at the magic ellipticity $\chi_{\text{m}}$ (star), the differential light shift with respect to $N=0$ is zero for one $N=1$ sublevel.}
    \label{fig:intro}
\end{figure}

Many approaches have been developed to reduce the differential light shift between rotational ground ($N=0$) and excited ($N\ge1$) states. These approaches include selecting a specific angle between the confining light's linear polarization and static magnetic or electric fields~\cite{Kotochigova2010, Neyenhuis_anisotropic12,seesselberg_extending18,blackmore_controlling_2020}, using a particular trapping wavelength~\cite{bause_tuneout20,guan_magic21, kondov_molecular_2019} or  intensity~\cite{blackmore_coherent_2020}, or a specific magnetic field~\cite{christakis_probing_2023}.
In the first approach, the  static field determines the orientation of the excited rotational eigenstates, and a specific polarization angle matches the polarizability of one  excited state to the ground state. 
This method, however, is not applicable even at moderate trap depths when the differential light shifts are of similar magnitude to the shifts induced by the static fields, such as for polar molecules confined in optical tweezers (Fig.~\ref{fig:intro} (a)).
In this deep trap regime, the rotational eigenstates are  determined by the polarization of the tweezer light, rather than an external static field. For open-shell ground-state $X^{2}\Sigma^+$ molecules such as CaF, 
an isotropic $F=0$ state within the $N=1$ manifold is available that eliminates the first-order differential light shift~\cite{Burchesky2021}, enabling observation of dipolar interactions~\cite{holland_-demand_2022, bao_dipolar_2022}. However, for other choices of state pairs, including those with the largest transition dipole moments, a  large first-order differential light shift is expected. 

In this Letter, we employ a method to trap $X^1\Sigma^+$ NaCs molecules in optical tweezers with ``magic'' elliptical polarization, to reduce the differential light shift 
by more than three orders of magnitude. Here, ``magic'' refers to a specific degree of ellipticity near $\chi_{\text{m}}=\frac{1}{2}\cos^{-1}(1/3)\approx35.26\degree$ that nulls the differential light shift~\cite{Rosenband2018}. Similar methods  have been explored in atomic systems~\cite{taichenachev_optical06,kim_magic13,cooper_alkaline18,trautmann_1s023}. We measure the reduction of differential shift by  microwave spectroscopy and use Ramsey interferometry to characterize the coherence. With the aid of dynamical decoupling pulses, we achieve a coherence time of 250(40)~ms.

Some theoretical aspects of ``magic'' ellipticity trapping have been described in Ref. \cite{Rosenband2018}. The ground state of rotation ($N=0$) is illustrated in Fig.~\ref{fig:intro} (a) as a spherically symmetric rotational wavefunction with isotropic polarizability $(2\alpha_{\perp}+\alpha_{||})/3$ for any optical polarization, where ($\alpha_{||}$) and ($\alpha_{\perp}$) are the molecule's parallel and perpendicular polarizability with respect to the internuclear axis.  This approximation is valid when the trap depth is small compared to the energy of $N=2$ excited states (the optical potential couples states with both $\Delta N=0$ and $\Delta N=2$)~\cite{supplemental}.  Throughout the text, \emph{trap depth} ($U$) refers to the optical potential experienced by the relatively unperturbed $N=0$ state.  This spectroscopic study uses frequency units that are implicitly related to energy by Planck's constant.

As shown in Fig.~\ref{fig:intro} (b), for $N=1$ the trap-induced light shift lifts the degeneracy of the three rotational sublevels ($m_N=-1,0,1$), and strongly perturbs their wavefunctions, such that each sublevel has well-defined orientation relative to the optical polarization above a certain trap depth threshold.  Due to molecular hyperfine structure and anisotropy of polarizability, this threshold is about 100~kHz for NaCs. In a linearly polarized trap the light shift is as large as 400.8 kHz/(MHz trap depth), or a ratio to trap depth of 0.4. By tuning the ellipticity~\cite{born_wolff_1980} near $\chi_{\text{m}}$, we can eliminate this differential light shift to first order.

We implement the magic ellipticity trapping scheme with an array of individual \Na\Cs{} molecules  in optical tweezers  prepared using  methods and an apparatus described previously~\cite{zhang_forming_2020, zhang_optical_2022}, with minor modifications described here.
In brief, we first load parallel tweezer arrays of individual \Na{} and \Cs{} atoms. The wavelengths of the trapping lasers are 616 nm for \Na{} and 1064 nm and for \Cs{}, and the spacing between neighboring traps is $\sim$5 $\mu$m.

The stochastically loaded atoms are then rearranged to a densely filled array of 8 traps for each species~\cite{Endres2016, Barredo2016}. After motional ground state cooling and state preparation \cite{yu_motional-ground-state_2018, liu_molecular_2019}, the Na atoms are adiabatically transported into the 1064 nm traps, and the atom pairs are converted into weakly bound molecules by sweeping the magnetic field across a Feshbach resonance~\cite{zhang_forming_2020} before holding at 864.5~Gauss. Subsequently, molecules are coherently transferred to their $X^{1}\Sigma^+$ rovibronic ground state with hyperfine quantum numbers $\ket{I_{Na},M_{Na},I_{Cs},M_{Cs}}=$ $\ket{3/2,3/2,7/2,5/2}$ via stimulated Raman adiabatic passage~\cite{Picard2023}. After molecule creation, we apply a pulse resonant with the \Cs{} $D_2$ transition to blast away any residual atoms. To detect molecules, we reverse the steps and image the atoms. The \Cs{} blast step provides a background free molecule signal.

\begin{figure}[t]    \centering\includegraphics{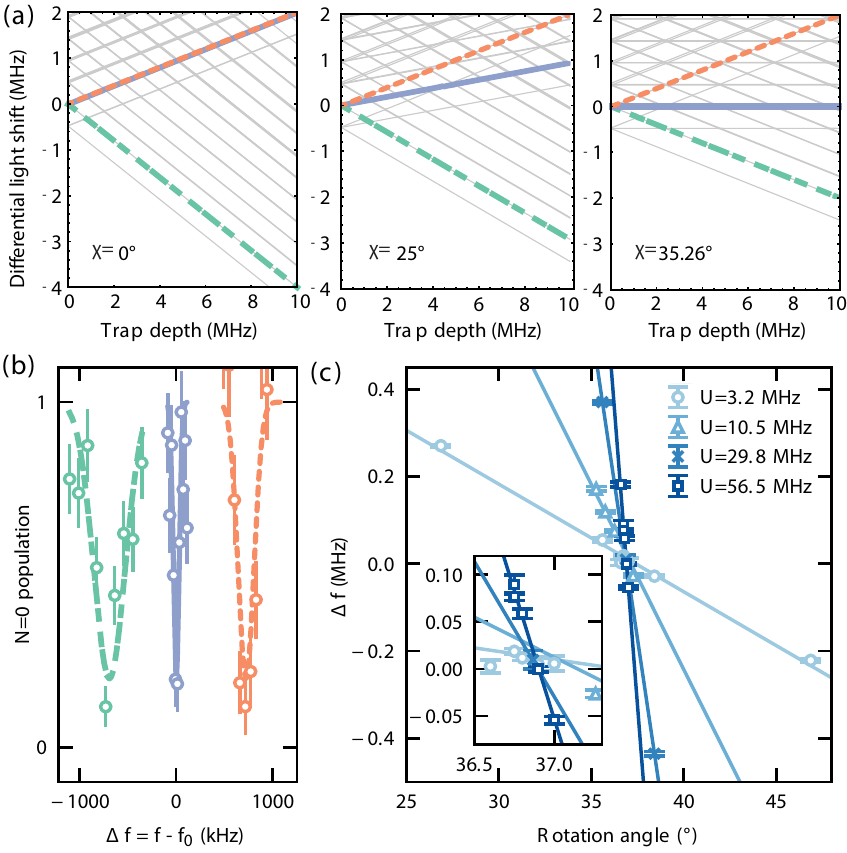}
    \caption{$N=0$ to $N=1$ rotational transition in the vibronic ground state of \Na\Cs{}. (a) Differential light shift as a function of $N=0$ trap depth at three distinct trap ellipticities. The colored red, blue, and green lines correspond to the light shifts of $N=1, m_N$ sublevels of the hyperfine state $\ket{3/2,3/2,7/2,5/2}$, and the grey lines correspond to the same for other hyperfine states. (b) An example of a measured rotational spectrum in the magic elliptically polarized trap.
    (c) Extracted resonant frequencies of the middle state from the rotational spectroscopy at different trap depths ($U$) and QWP rotational angles. The solid lines are linear fits.
    }
    \label{fig:spectroscopy}
\end{figure}

Because atomic state preparation, cooling, and detection require linearly polarized tweezer light, it is necessary to change the polarization from linear to elliptical and back during the experiment sequence. 
For this purpose, a 
motorized stage (Griffin Motion, RTS100) rotates a quarter-wave plate (QWP) by $\chi_m$ in about 100 ms with a repeatability of $\pm$0.0007$^{\circ}$. To ensure polarization purity and minimize  site-to-site polarization variation across the array, we  use a Glan-Taylor polarizer and place the QWP as the last element before the microscope objective.  Before the QWP, the polarization extinction ratio is approximately 300,000.

To characterize differential light shifts under various trap polarizations and intensities, we perform rotational microwave spectroscopy to selectively transfer the molecules from $N=0$ to the relevant $N=1$ sublevel with a transition frequency near 3.47~GHz. The microwave pulses are generated by a tunable source referenced to a stable Rubidium clock. As trap ellipticity increases (Fig.~\ref{fig:spectroscopy}a), the degeneracy of the two upper sublevels is lifted.  At an ellipticity near $\chi_m$ the state with no differential light shift emerges. 
An example of the $N=0$ to $N=1$ microwave spectrum in an elliptical trap is shown in Fig.~\ref{fig:spectroscopy}b.

To find the precise QWP angle that achieves magic ellipticity, we
scan the microwave frequency over the transition with a 10 $\mu$s  $\pi$-pulse at varying trap depths and record the resonance frequencies~(Fig.~\ref{fig:spectroscopy}c). As expected, the slope of the resonance frequency as a function of QWP rotation is steeper at higher trap depths. The differential light shift is zero where the rotation angle dependence for different trap depths intersect. We determine the angle of the intersection with a weighted fit to be 36.83(10)$^{\circ}$, which deviates from the theoretically expected magic ellipticity angle by 1.57$^{\circ}$ ~\cite{supplemental}. The discrepancy may be due to birefringence of the glass cell assembly and  the microscope objective. Nulling the differential lights shift allows a determination of the $N=0$ to $N=1$ transition frequency $f_0=3.4713203(7)$~GHz, taken as the transition frequency at the optimal ellipticity at trap depth $U=$1.34~MHz (see \cite{supplemental} for trap depth calibration) and is consistent with the low-depth regime measurement of the transition in Ref. \cite{bigagli2023collisionally}.

\begin{figure}[t]
    \centering \includegraphics{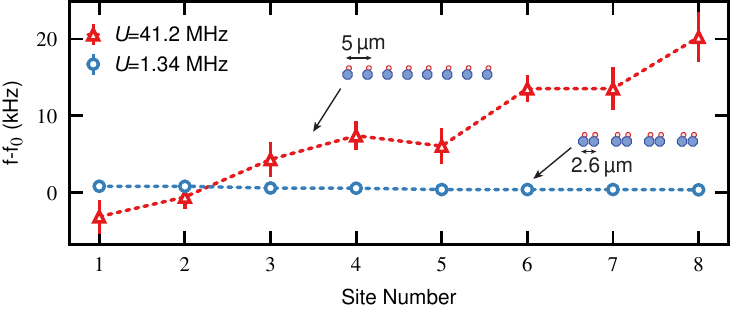}
    \caption{Variations in rotational transition frequency across the 8 traps at two different trap depths $U$, and trap geometries and spacing. At a high trap depth of 41.2 MHz (red triangles) the transition frequency spans a range of 23 kHz across the array and at a lower depth of 1.34 MHz (blue circles) it spans $< 1$ kHz.}
    \label{fig:twobody}
\end{figure}

Residual light shift at the optimal QWP angle reveal site-to-site variations in frequency across the 8 trap sites.  To characterize these effects, we use a 60~$\mu$s microwave pulse to drive the rotational transition at a large trap depth of 41.2~MHz that magnifies light shifts. We find a site-to-site variation spanning 23~kHz (Fig.~\ref{fig:twobody}) with a 5(2)~kHz average shift from the measured $f_0$, which constitutes a light shift to trap depth ratio of $1.2(5)\cdot 10^{-4}$.  Despite the variation, this corresponds to a reduction in sensitivity by three orders of magnitude compared to the linearly polarized trap. The trap intensities across the array are made uniform to within $1$\%, such that they do not significantly contribute. We attribute the residual shifts to an ellipticity variation of 0.062$^{\circ}$ across the array. For the aspects of polarization ellipticity and intensity considered here, we expect negligible decoherence when a spin-echo pulse removes static frequency shifts.

\begin{figure}[t]
    \centering
    \includegraphics{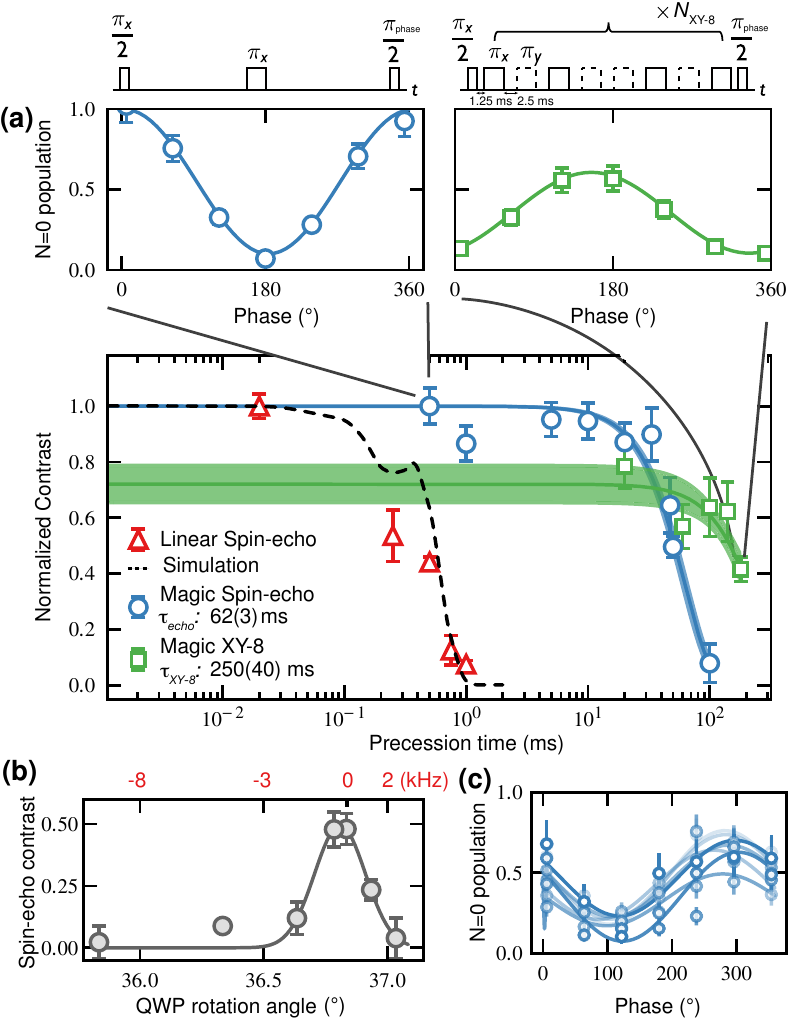}
    \caption{Rotational coherence times for approximately 1.3~MHz trap depth. (a) 
    The coherence time is characterized using spin-echo phase Ramsey pulse sequence (shown in top left) in linear and magic elliptically polarized traps.
    In the linear trap, the phase Ramsey contrasts as a function of the free precession time (red, triangle) agree well with the simulated coherence decay based on intensity noise (black dotted line).
    The spin-echo coherence time is extended by two orders of magnitude in the magic trap (blue circle), which can be further improved (green square) using the XY-8 pulse sequence illustrated in the top right. The solid lines represent the fit to a Gaussian model $C(t)=\mathrm{exp}[-(t/\tau)^2]$. For the spin echo data, all contrasts are normalized to the shortest time point. For the XY-8 pulse sequence,  the overall amplitude is an additional fit parameter. An example spin-echo and XY-8 phase Ramsey scans are shown in the top left and right insets, respectively.  The contrast in the insets and for the XY-8 scan are normalized to the N=0 population.
    (b) Spin-echo phase Ramsey contrast at 50 ms as a function of the QWP rotation angle. A Gaussian fit yields an optimal rotation angle of 36.81(3) degrees. The top axis is the corresponding light shifts expected from the ellipticity angle deviation relative to $f_0$.  (c) Site-by-site Ramsey contrasts at 50 ms showing a global phase shift of 73(3)$\degree$. The contrast is normalized to the averaged N=0 population.} 
    \label{fig:ramsey}
\end{figure}

With the reduced light shift sensitivity, $N=0$ and $N=1$ rotational superpositions exhibit long coherence that we characterize via Ramsey spectroscopy. Although the ensemble-averaged contrast would decay rapidly due to static light shift variation across the traps~\cite{supplemental}, a spin-echo $\pi$-pulse eliminates such dephasing. For a linearly polarized trap ($U=1.0$~MHz), the $1/e$ decay time is $\tau=0.57(2)$ ms, in agreement with a simulated coherence decay that incorporates measured intensity noise and the strong light shift sensitivity (Fig.~\ref{fig:ramsey}a black line)~\cite{supplemental}.
With optimal ellipticity, the spin-echo coherence is extended by two orders of magnitude to 62(3)~ms (blue circles in Fig.~\ref{fig:ramsey}a). This coherence is further extended to 250(40)~ms by use of repeated XY-8 sequences (up to 72 total pulses)~\cite{GULLION1990,li_tunable23}.  Figure~\ref{fig:ramsey}b shows that the observed  spin-echo coherence contrast depends sensitively on small changes of the QWP angle in the vicinity of magic ellipticity. The spin-echo contrast for individual traps at a precession time of 50 ms (Fig.~\ref{fig:ramsey}c) shows a small amount of dephasing between sites.  An overall phase shift of $73(3)^{\circ}$ at long times (despite spin-echo) indicates a changing global frequency whose source is uncertain.

 The effects of global intensity noise were simulated and do not account for observed decoherence~\cite{supplemental}.  Magnetic field fluctuation is also unlikely to be the cause, as the transition sensitivity is below 2~Hz/G while the field noise amplitude is $10^{-2}$~G.  However, electric fields of $0.5$~V/m with fluctuations of $0.012$~V/m were measured in a similar vacuum glass cell environment~\cite{ocola2022control}. Due to the large electric dipole moment of NaCs (4.6~Debye) and its easily polarizable nature, the quadratic Stark shift at such a field would cause frequency fluctuation of up to 12.6 Hz, which may explain the decoherence (see~\cite{supplemental} for a Monte Carlo simulation).

Beyond single-body decoherence, a natural question is whether dipolar interaction causes the observed decoherence.  However, because the polarization ellipse determines the dipolar axis, the interaction is reduced to zero in the present geometry $\phi=\chi_m$, with molecule separation along $\mathbf{x}$ (Fig.~\ref{fig:intro}). We verify this experimentally by dropping molecules from every other site to extend the distance of the neighboring molecules to 10~$\mu$m and observe no change in the coherence. In the future, a half-wave plate will allow adjustment of the anisotropic dipolar interaction into a maximal head-to-tail configuration while maintaining the magic condition.
In conclusion, we have demonstrated magic elliptical polarization trapping of polar molecules in the deep trap regime. The method reduces the light shift sensitivity between particular sublevels of the lowest two rotational states by three orders of magnitude and achieves a spin-echo rotational coherence time of 62(3) ms.  This exceeds the expected 2~ms duration of dipolar entangling gates by a factor 30 for 2.6 $\mu$m molecule spacing (see Fig.~\ref{fig:twobody}). Coherence, limited by slow drifts that potentially arise from electric field fluctuations, may be further extended by apparatus improvements and dynamical decoupling.
Additional tunable control over molecular dipole orientation will bring coherent dipolar interaction between motional ground-state molecules in tweezers within reach, leveraging the rich properties of molecules to enable high-fidelity gates~\cite{ni_dipolar_2018}, simulation of exotic phases~\cite{Yao2018, homeier2023antiferromagnetic}, and state engineering~\cite{Sundar2018}.

\begin{acknowledgements}
  We thank Yi-Xiang Liu, Fang Fang, and Markus Greiner for  discussions. This work is supported by AFOSR (FA9550-19-1-0089), NSF (PHY-2110225), and AFOSR-MURI (FA9550-20-1-0323).   

\end{acknowledgements}
\textit{Note added -} A related work demonstrates ``magic'' wavelength trapping of  polar molecules~\cite{gregory2023secondscale}.

\bibliography{magic_ell,master_ref_June2023}

\end{document}


\title{Supplemental Material: Extended rotational coherence of polar molecules in an elliptically polarized trap}

\author{Annie J. Park}
\thanks{A.J.P. and L.R.B.P contributed equally to this work.}
\affiliation{\HarvardChemistryaddress} 
\affiliation{\HarvardPhysicsAddress}
\affiliation{\CUAaddress}

\author{Lewis R.B. Picard}
\thanks{A.J.P. and L.R.B.P contributed equally to this work.}
\affiliation{\HarvardPhysicsAddress}
\affiliation{\HarvardChemistryaddress} 
\affiliation{\CUAaddress}

\author{Gabriel E. Patenotte}
\affiliation{\HarvardPhysicsAddress}
\affiliation{\HarvardChemistryaddress} 
\affiliation{\CUAaddress}

\author{Jessie T. Zhang}
\altaffiliation[Current affiliation: ]{University of California, Berkeley}
\affiliation{\HarvardChemistryaddress} 
\affiliation{\HarvardPhysicsAddress}
\affiliation{\CUAaddress}

\author{Till Rosenband}
\affiliation{Agendile LLC, Cambridge, Massachusetts, 02140, USA}

\author{Kang-Kuen Ni}
\email{ni@chemistry.harvard.edu}
\affiliation{\HarvardChemistryaddress} 
\affiliation{\HarvardPhysicsAddress}
\affiliation{\CUAaddress}

\date{\today}

\maketitle

\section{Trap depth calibration}
We calibrate our molecule trap depth using an experimental measurement of the trap frequency of Cs atoms and scale by a theoretical value of the NaCs polarizability from \cite{Vexiau2017}. Our trap intensity is controlled by an analog intensity servo which monitors a reflection picked off from a wedge in the tweezer beam path. We use an independent measurement of the beam power at an additional picked-off location to precisely calibrate the voltage setpoint of the servo and correct for a small 0.05 V offset of the zero-point of the servo output. At a servo setpoint of 3.4 V, just below our maximum achievable trap depth, we measure a Cs trap frequency of 146.5(1) kHz, corresponding to a trap depth of 108.4(6) MHz \cite{Grimm2000a}, using the Cs polarizability of 1163.4 au at 1064 nm. We then use the theoretical NaCs polarizability of 939.8 to infer a molecule trap depth of 83.5(5) MHz at this intensity setpoint. All other trap depths in the main text are scaled from this value using the servo setpoint.

\section{Optical Layout and Waveplate calibration}
\begin{figure}[h]    \centering\includegraphics[width = 0.6\textwidth]{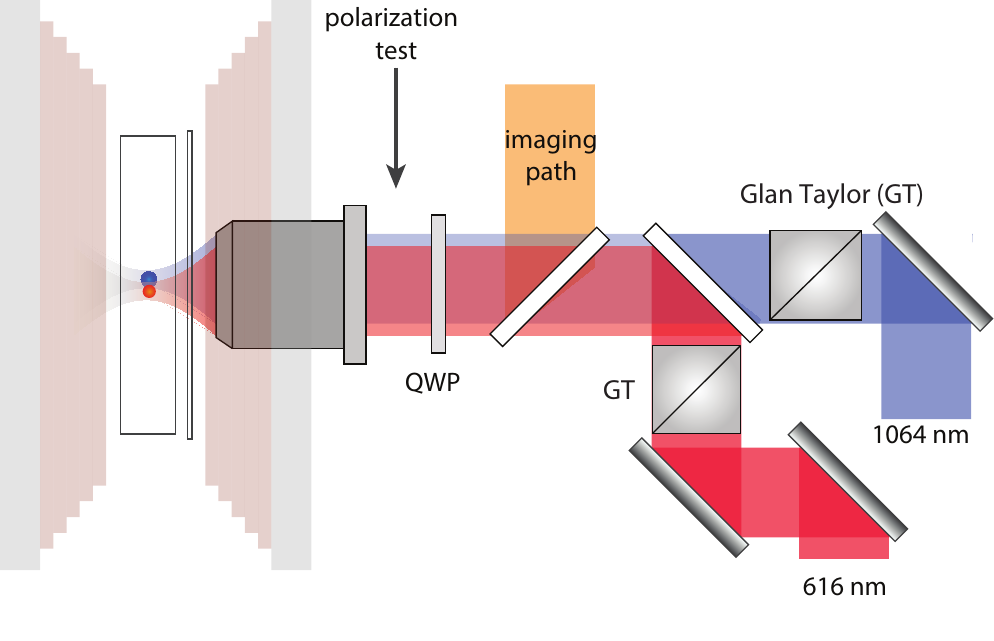}
    \caption{Schematic of the optical path (adapted from~\cite{zhang_assembling_2021}). The polarizations of the 616 nm and 1064 nm tweezer beams are individually cleaned up using Glan-Taylor polarizers before they are combined using dichroic mirrors. After the dichroics, the beams pass through a waveplate, that is quarter (QWP) for 1064 nm and half for 616 nm. The polarization purity of 1064 nm beam was measured using a thin-film polarizer after the QWP.}
    \label{fig:beampath}
\end{figure}
The optical path to trap and image the molecules is shown in Fig. S\ref{fig:beampath}. We initialize each tweezer with linear polarization using an anti-reflection coated Glan-Taylor polarizer. The tweezers are combined on a dichroic whose vertical tilt is tuned to preserve linear polarization with an extinction ratio of $\sim3\cdot 10^5$ for 1064 nm. The beam then passes through a waveplate in a motorized rotation mount with quarter wavelength retardance for 1064 nm and half wavelength retardance for 616 nm. The ellipticity and orientation of each beam's polarization was measured using a thin-film polarizer in a manual rotation mount placed between the objective and waveplate. Specifically, the ellipticity was measured to be $\chi=\tan^{-1}(\sqrt{\frac{P_\text{min}}{P_\text{max}}})$ where $P$ is the power of minimum or maximum transmitted power as a function of polarizer angle.

\section{Adiabatic Crossings and Second-order shifts}

\begin{figure}[h]    \centering\includegraphics[width = 0.6\textwidth]{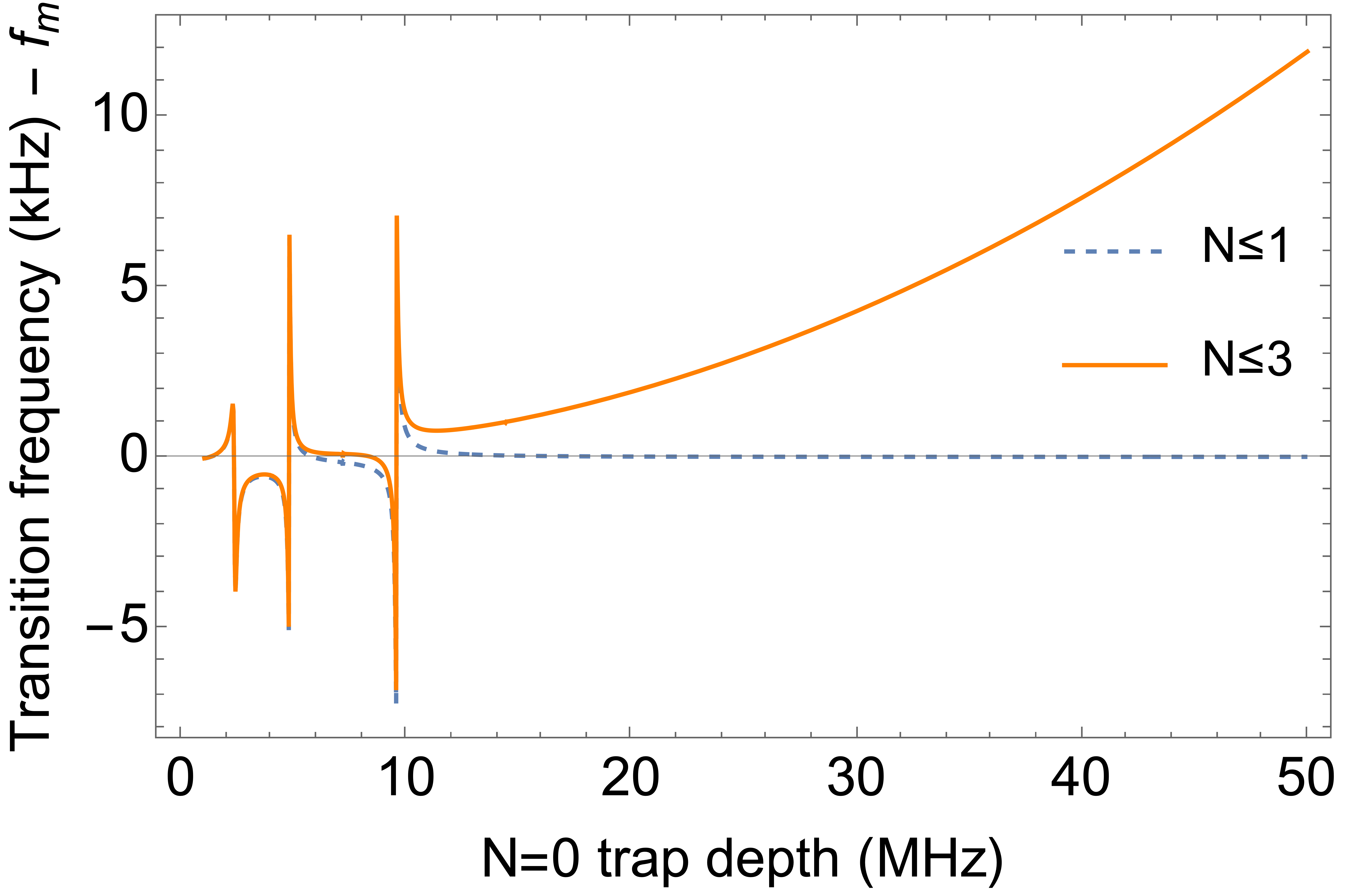}
    \caption{Calculation of the transition frequency to the magic $N=1$ state ($f_m$). The effective Hamiltonian for the rotational/hyperfine manifold of NaCs is diagonalized with a basis extending up to either $N=1$ or $N=3$. The difference represents the second order Stark shift, or 'hyperpolarizability', which mixes $N=0$ with $N=2$ and $N=1$ with $N=3$. The three most prominent adiabatic crossings are produced by the coupling of high field seeking $N=1$ hyperfine states $|\frac{1}{2},\frac{7}{2}\rangle, |\frac{3}{2},\frac{1}{2}\rangle, |\text{-}\frac{1}{2},\frac{5}{2}\rangle$ and the low field seeking hyperfine state $|\frac{3}{2},\frac{7}{2}\rangle$ with the magic $|\frac{3}{2},\frac{5}{2}\rangle$ state.}
    \label{fig:n3Correction}
\end{figure}

\begin{figure}[h]    \centering\includegraphics[width = 0.6\textwidth]{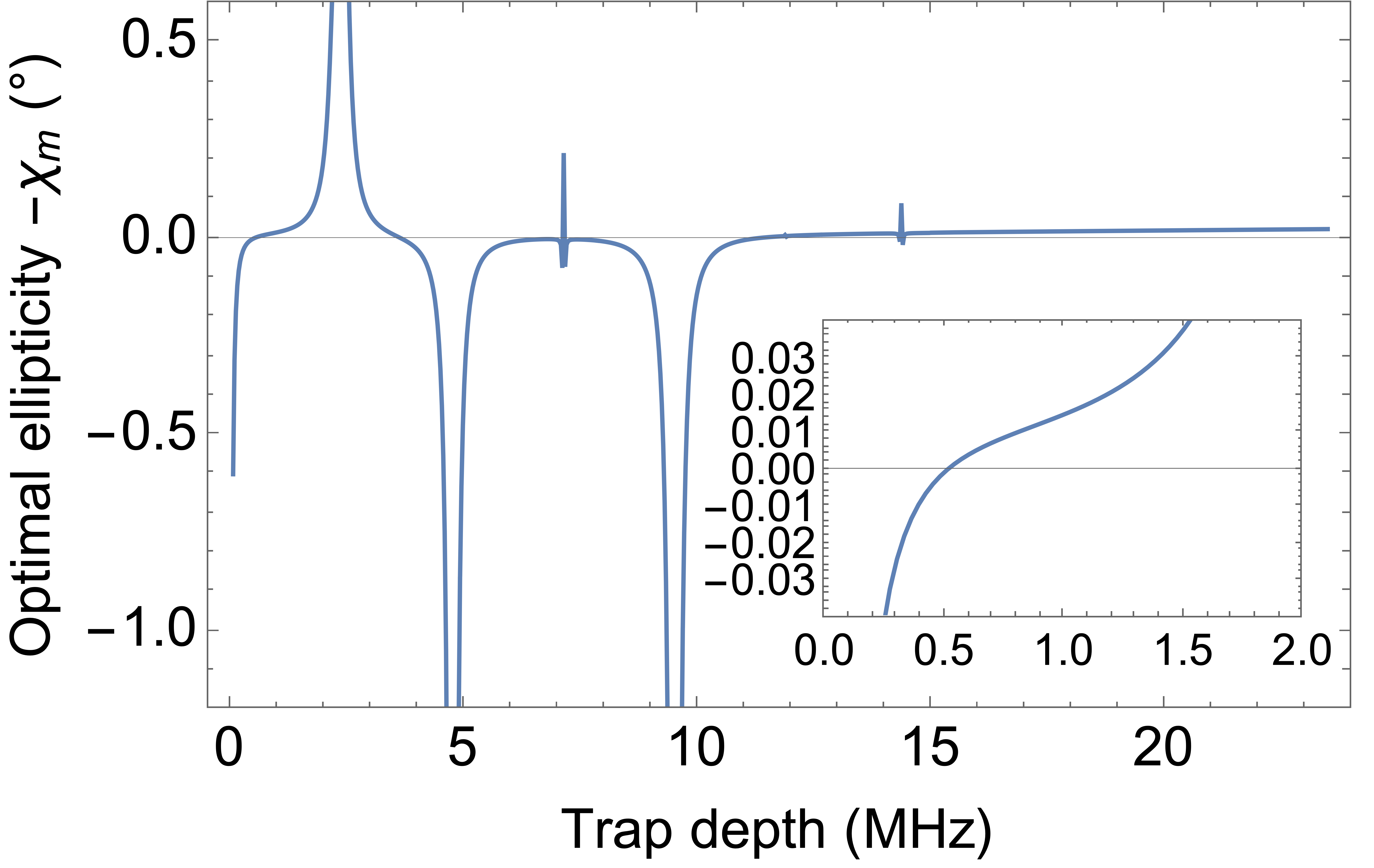}
    \caption{Ellipticity relative to $\chi_m$ for which the transition frequency is first order insensitive to intensity fluctuations. The inset highlights the optimal ellipticity for the range of trap depths used in coherence time measurements.}
    \label{fig:optimalEllipticity}
\end{figure}

We diagonalize the effective Hamiltonian for the rotational-hyperfine states of NaCs given in \cite{Aldegunde2009} and find two effects which make the theoretical magic state somewhat first-order sensitive to light shifts for most tweezer intensities in the ``deep'' trap regime. The first effect is the presence of adiabatic crossings of the magic N=1 state with other rotational hyperfine states at certain trap depths. These crossings are shown on the left side of Fig.~S\ref{fig:n3Correction} and occur when an electric quadrupole coupling between rotational hyperfine states of order $ \sim 10^{-1}\,eQq\approx10$ kHz is made resonant. Resonances occur due to the unequal polarizability of the $N=1$ rotational states and a Zeeman shift between hyperfine states. These crossings introduce a first order light shift to all trap depths, with a magnitude that depends on the depth. At a trap depth of 1.34 MHz, this results in a first order sensitivity of $269$ Hz/MHz. The first order light shift can be eliminated by changing to a unique near-magic ellipticity for each depth, shown in Fig.~S\ref{fig:optimalEllipticity}, which is 0.025\degree~ larger than $\chi_m$ at a trap depth of 1.34 MHz. This depth-corrected ellipticity is the angle to which our empirical calibrations of the rotation angle in the main text Figs. 2(c) and 4(b) are sensitive.

The optical tweezer also produces a second order light shift by coupling rotational states separated by $\Delta N=2$. The higher rotational states that are mixed in to the $N=0$ and $N=1$ states are anisotropic, which breaks the magic condition. The transition energy increases quadratically with intensity, which is shown on the right side of  Fig.~S\ref{fig:n3Correction}.

\section{Rotational decoherence due to tweezer intensity noise}

We measure the tweezer light intensity on an out-of-loop photodiode. The signal is bandpassed filtered and amplified through a low-noise SR560 preamplifier from Stanford Research Systems. From the noise spectra of the tweezer light intensity, we can simulate the decay of coherence contrast using the Magnus expansion then the cumulant expansion. Details on the method can be found in \cite{ocw_22_51_notes} and references therein. Specifically, we consider only the second cumulant, which gives the highest order contribution to coherence decay. This is expressed as
\begin{equation}
    c_2(t) = \frac{1}{2}\int_{-\infty}^\infty d\omega S(\omega)|F(\omega t)|^2
\end{equation}
where $S(\omega)$ is the power spectrum of the noise source, and $F(\omega t)$ is the transfer function of the pulse sequence.
Then the contrast follows
\begin{equation}
    p = \exp(-c_2(t))
\end{equation}
We assume the $\pi$-pulses are delta-pulses, so that the transfer function for spin echo is
\begin{equation}
    F(\omega, t) = \frac{4}{\omega}\sin\left(\frac{\omega T}{4}\right)^2
\end{equation}
We use this method to simulate decoherence due to intensity noise for molecules in a linearly polarized trap. The simulation curve is shown in Fig.4(a) and is in good agreement with the observed coherence time.

With the three orders of magnitude reduction in sensitivity to intensity fluctuations achieved using magic ellipticity, the dominant contribution to the decay of coherence will be from low frequency noise on the Hz scale. Slow drifts on this timescale can give rise to overall phase shifts in the spin echo Ramsey signal, which may vary shot-to-shot, reducing the contrast of the time-averaged signal. Because of the gradient of ellipticities across sites in the array there will also be a different phase shift accumulated for each site, which will further impact the averaged signal. 

\begin{figure}[h]    \centering\includegraphics[width = 0.9\textwidth]{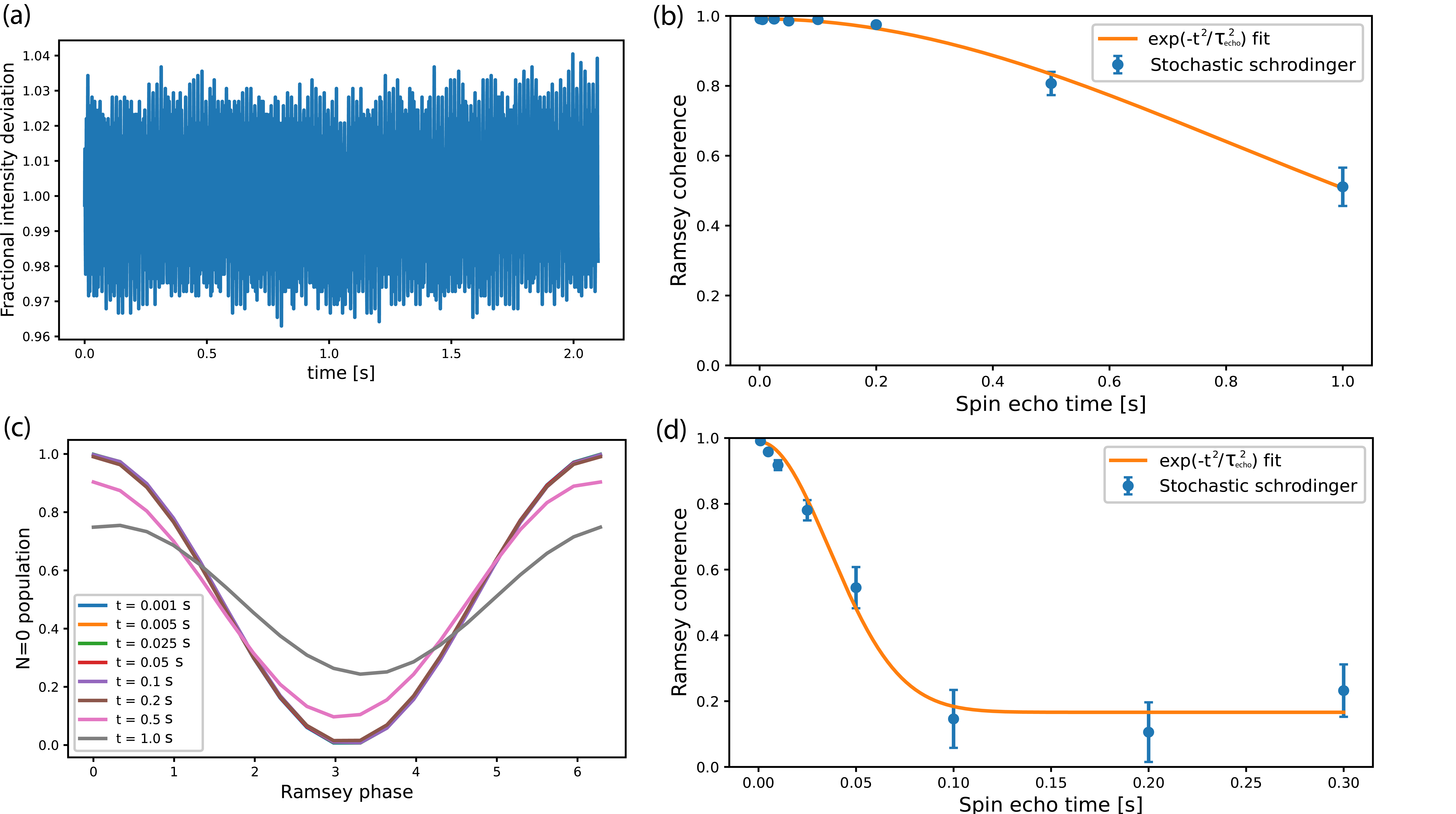}
    \caption{(a) Measured intensity fluctuation of tweezer beam over 1.8 s. (b) Normalized Ramsey coherence for average of 8 sites predicted from Monte Carlo simulation using random sampling of intensity noise data and experimentally calibrated light shifts of each site. Fitting to a Gaussian decay gives a coherence time of  $\tau_{\mathrm{echo}} = 1.24$ s. (c) Corresponding simulated Ramsey phase scans for each time point in b, showing both decay of coherence and slight overall phase shift due to slow drifts. (d) Simulated Ramsey coherence assuming an additional random Gaussian detuning noise source with a standard deviation of $2\pi \times 25$ Hz. Fitting to this data yields a decay time of $\tau_{\mathrm{echo}} = 59$ ms. }
    \label{fig:spinEchoSim}
\end{figure}

To model these effects, we measure the slow drift of our tweezer intensity using an out-of-loop photodiode with no frequency filtering or amplification for a time of 5 s, representing approximately between 3 and 4 experimental cycles. We then model the Schr\"odinger evolution of the rotational states of all 8 molecule sites under a spin-echo pulse sequence with a time-varying detuning for each site sampled directly from the intensity fluctuation data, multiplied by the average light shift measured in the main text Fig. 3. To capture the effect of shot-to-shot variations, we determine the coherence at each time using an ensemble average of 20 runs with randomly chosen $t = 0$ startpoints within the intensity noise trace. Using this method, we find an expected coherence time which is well described by a Gaussian  $e^{-t^2/\tau_{\mathrm{echo}}^2}$ decay with a characteristic timescale of $\tau_{\mathrm{echo}} = 1.24$ s, more than an order of magnitude longer than the spin echo decay time we observe experimentally. This strongly suggests that we are limited by a noise source other than intensity fluctuations.

In order to better understand potential additional noise sources, including electric field noise as described below and in the main text, we model our expected coherence in the presence of a random Gaussian noise source sampled with a bandwidth of 1 kHz. Using this method we estimate that detuning noise with a standard deviation on the order of $2\pi \times 25$ Hz would explain our observed spin echo coherence time, as shown in Fig. S\ref{fig:spinEchoSim}(d). To simulate slow drifts in detuning which could give rise to the overall phase shift of the spin echo signal shown in the main text Fig. 4, we assume a detuning that varies linearly in time. We find that to explain our observed phase shift the rate of change of this detuning would need to be at least $\frac{d\Delta}{dt} = 2\pi\times330$ Hz / s, corresponding to a total change in detuning of $2\pi\times16.5$ Hz over the course of the 50 ms spin echo wait time.

\section{Electric field sensitivity}

\begin{figure}[h]    \centering\includegraphics[width = 0.5\textwidth]{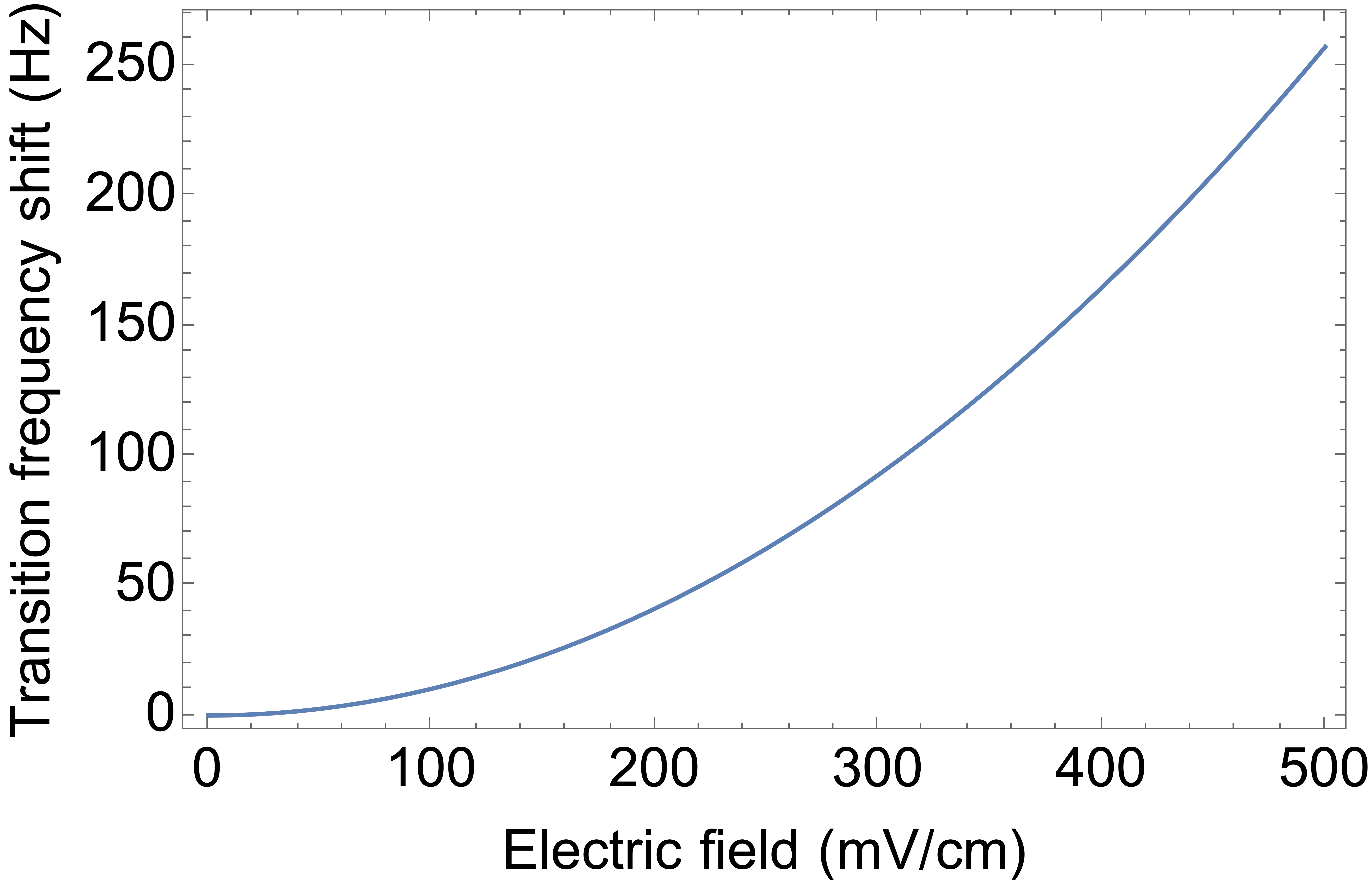}
    \caption{Shift of transition frequency from $f_m$ at a trap depth of 1.34 MHz as a function of applied electric field oriented parallel to the minor axis of the polarization ellipse.}
    \label{fig:EField}
\end{figure}

We calculate the expected shift in the transition frequency from $N=0$ to the $N=1$ magic state as a function of applied electric field oriented parallel to the major axis of the polarization ellipse, which is the orientation which gives the maximum differential shift. The dependence on the field is quadratic, with a fitted dependence of $f=0.1~E^2$ Hz / (V m$^{-1}$)$^2$. If we assume a background static electric field of 500 mV/cm, then a fluctuating component with a standard deviation of 25 mV/cm would be sufficient to generate the $2\pi \times 25$ Hz detuning noise required to explain our spin-echo observed coherence time, which is of a similar order to that measured in Ref ~\cite{ocola2022control}.

\section{Ellipticity variation sampled by a single molecule in a focused tweezer}

\begin{figure}[h]    \centering\includegraphics[width =\textwidth]{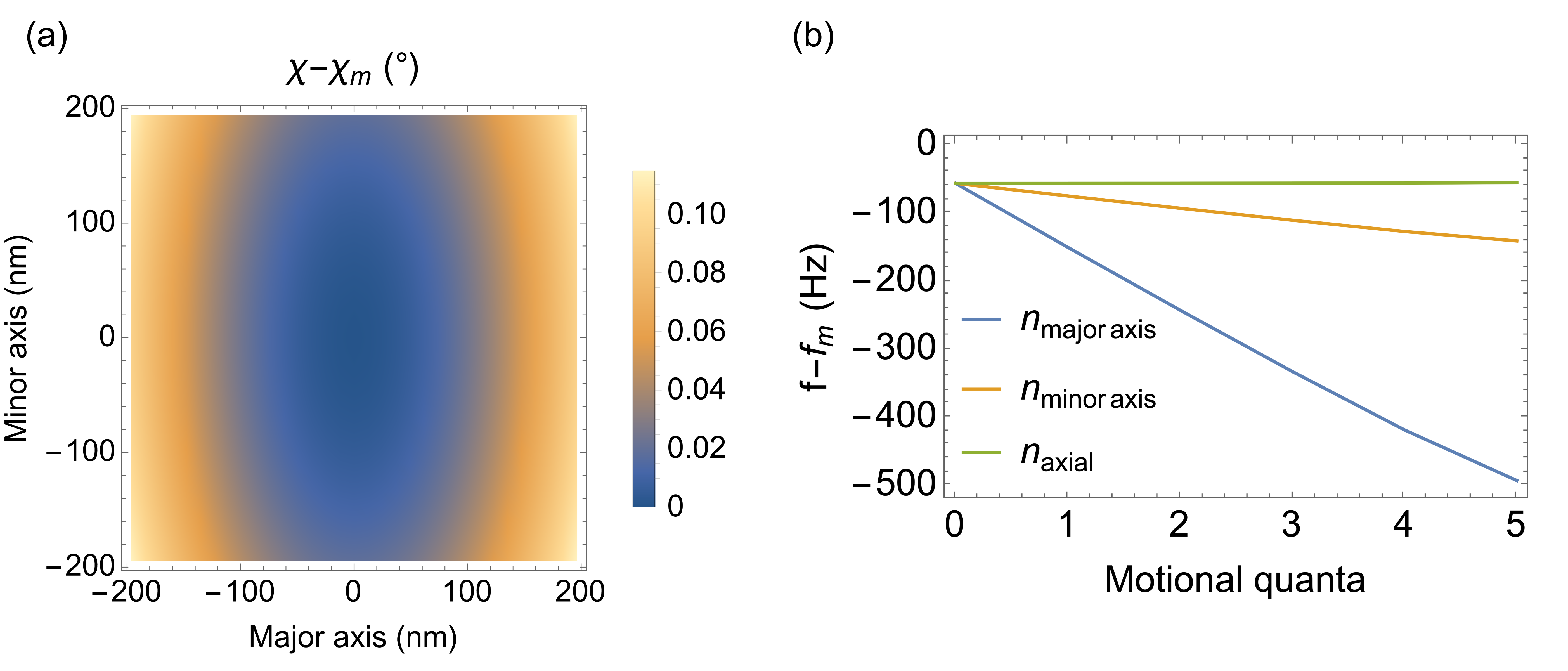}
   \caption{(a) Ellipticity  relative to the magic ellipticity of 35.26$\degree$  in the focal plane of a 1064 nm optical tweezer. (b) Shift in the transition frequency of a molecule in an optical tweezer due to motional excitation, which allows the molecule to sample regions of higher ellipticity. Motional excitations along the major axis result in the largest detuning, since axially polarized light is concentrated along the major axis. It is assumed in this calculation that the ellipticity at the focus of the trap is the optimal $\chi_m$.}
    \label{fig:ellipticityFocalPlane}
\end{figure}

We use vector Debye theory \cite{Chen2009_tightfocus} to estimate the ellipticity variation experienced by a molecule trapped in an optical tweezer. The tight focusing of light to a beam waist close to $\lambda$ results in axial polarization localized on the edge of the beam that alters both the orientation and ellipticity of the tweezer polarization. The power transferred to the axial direction comes mainly from the major axis of the polarization ellipse, so a molecule experiences a greater ellipticity the further it is from the focus. The ground motional wavepacket at a trap depth of 1.34 MHz has a full width at half maximum of 98.3 nm in the major axis direction, 97.3 nm in the minor axis direction, and 225.2 nm in the axial direction.  In Fig.~S \ref{fig:ellipticityFocalPlane}, we compute the ellipticity distribution at the focus and derive the resulting shift in transition frequency for a motional wave packet with excitation in the radial and axial directions. 

\bibliography{magic_ell,master_ref_June2023}